\documentclass[%
 reprint,
bibnotes,
 amsmath,amssymb,
 aps,
floatfix,
]{revtex4-2}

\usepackage{graphicx}
\allowdisplaybreaks

\usepackage{xcolor}

\begin{document}

\title{Inferring Local Structure from Pairwise Correlations}

\author{Mahajabin Rahman}
\email{mahajabin.rahman@emory.edu}
\affiliation{Department of Physics, Emory University, Atlanta, GA 30322, USA}
\author{Ilya Nemenman}
\email{ilya.nemenman@emory.edu}
\affiliation{Department of Physics, Department of Biology, and Initiative in Theory and Modeling of Living Systems, Emory University, Atlanta, GA 30322, USA}%
\date{\today}

\begin{abstract}
To construct models of large, multivariate complex systems, such as those in biology, one needs to constrain which variables are allowed to interact. This can be viewed as detecting ``local'' structures among the variables. In the context of a simple toy model of 2D natural and synthetic images, we show that pairwise correlations between the variables---even when severely undersampled---provide enough information to recover local relations, including the dimensionality of the data, and to reconstruct arrangement of pixels in fully scrambled images.  This proves to be successful even though higher order interaction structures are present in our data. We build intuition behind the success, which we hope might contribute to modeling complex, multivariate systems and to explaining the success of modern attention-based machine learning approaches. 
\end{abstract}

\maketitle


{\em Introduction.}  The problem of data-driven inference of  laws governing a complex system has become a staple of what theorists do \cite{carleo2019machine}. This is particularly true in fields like biological physics, where high-throughput experiments generate large datasets, consisting of $T\gg1$ samples of $N\gg1$ measured variables, such as the activities of neurons or the presence or absence of mutations \cite{Savin2017,Marks2011}.  A probabilistic model $P(\mathbf{x})$ of such a system would include exponentially many coupling constants, which is unrealistic for experiments with $N \sim 10^2 \dots 10^3$ and $T \sim N \dots 10N$ at best. In traditional physical systems, the combinatorial problem does not exist because variables can only interact over short distances, so the total number of interaction coefficients is $O(N)$. The interaction structure may also be sparse even for complex biological systems (e.~g., a neuron may project into many, but not all other neurons). However, using sparsity is difficult until one knows which specific variables can interact \cite{Donoho2006, bailly-bechet2010, Ganguli2012, Bulso2016}.  Can we infer this effective local structure for complex systems from data alone? (Note that somewhat similar questions are also asked in the field of detecing community structure in complex networks  \cite{girvan2001, Clauset2005, Ravasz2003}.)

Physical locality and the constraints it imposes on the interactions are strongly dependent on the physical dimensionality of a problem. The effective dimensionality is usually either unknown or undefined for many complex biological systems. However, if any local, sparse interaction structure exists, it can be specified by a list  of interaction partners (effective neighbors) of each variable under consideration, requiring only $O(N)$ numbers. For comparison, the covariance matrix between the measured variables has $O(N^2)$ independent numbers, though many may be noisy. Therefore, it is plausible that the locality can be determined from the pairwise covariance matrix alone, even for systems where interactions can be of higher order or long-range (though presumably decaying with the effective distance between the interacting variables). This counting argument suggests that such locality reconstruction might be possible even when $T \sim N$, so that most empirical correlations are dominated by statistical noise \cite{potters2020first}. Explicitly showing that this can be done in a specific problem is the goal of this article.


We show that the structure of the pairwise covariance matrix is sufficient to reconstruct effective local relations, even when interactions are manifestly dense and of higher order. Specifically, we analyze a dataset of black and white images, where the notion of locality is clear, the dimensionality of the data is known (images are planar), correlations are critical and hence strong, and higher order interactions among pixels abound \cite{ruderman1994statistics}. We randomly permute the geometric location of the pixels and show that, based only on pairwise correlations within these randomized data, we can recover the dimensionality of the problem and restore the geometric arrangement of the pixels to a  high accuracy, reconstructing the original images with relatively small computational costs. This result is encouraging for using pairwise correlations to determine local interaction neighborhoods in more complex datasets where the true geometric structure is unknown, and it opens doors to investigating if modern neural networks implicitly perform similar computations.

\begin{figure*}
\includegraphics[width=140mm]{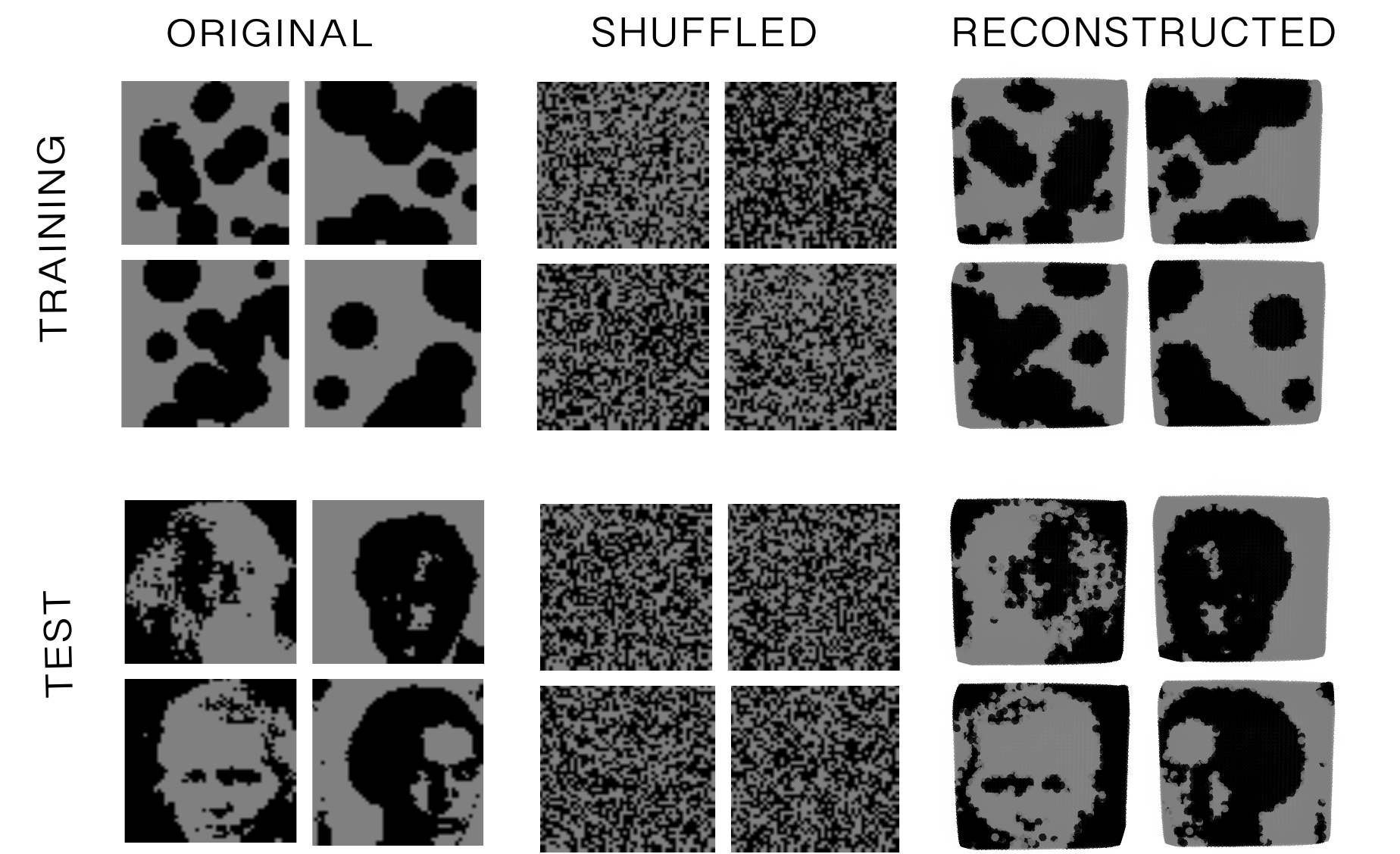}
\caption{\label{fig:summaryfig} \textbf{Original, shuffled, and reconstructed images.} The top half of the figure shows four examples from the {\em dead leaves} set of images (see text), which we use for training and detection of local pixel neighborhoods. The bottom half shows four examples of natural images, the test set, which we analyze based on local relations inferred for the training data. Pixels from original images (left) are shuffled according to one fixed random permutation (center). We then determine the optimal embedding of the shuffled pixels using  t-SNE, such that pixels highly correlated over the training set are embedded closer to each other (see text). The shuffled images are then reconstructed (right) by mapping their pixels into planar coordinates according to this optimal embedding, and then performing a single global rotation and rescaling of all images to closer match the originals. Qualitatively (see text for quantitative metrics), training and test reconstructed images are very close to the original ones, indicating that correlations can be used for defining local neighborhoods in, at least, image data.}
\end{figure*}

{\em Problem setup.}
In order to investigate if locality manifests itself in correlations, we require large, uniform datasets, and hence a generative model of images with features and correlations of different scales. Additionally, we need the dependence of pixel-pixel correlation on pixel separation to be generally similar to that of natural images. To accomplish this, we generate $T=40,000$ synthetic images as our training set, $I_i ({\bf x}), i=1,\dots, T$, ${\bf x}=(x,y)$, and  $0\leq(x,y)\leq 50$, using the dead leaves model \cite{Matheron1974, Lee2001, Pitkow2010}. This model simulates occlusion in natural images by piling opaque round objects onto an empty 50x50 canvas. We choose the maximum size of opaque circles, $r_{i,{\rm max}}$, at random from a uniform distribution between 1 and 19 pixels, and determine the number of circles of radius $r_{i,{\rm max}}$ required to make the image 50\% opaque, denoted as $n_i$. We then uniformly sample the centers of the $n_i$ circles and choose their radii uniformly between 0 and $r_{i,{\rm max}}$. The circles are placed on the canvas, and every pixel covered by at least one circle is marked as opaque, so that typically less than 50\% of all pixels in each image end up being opaque, as shown in Fig.~\ref{fig:summaryfig}. To demonstrate the weak sensitivity of our conclusions to these specific parameter choices, we also assemble a second dataset consisting of 5000 50x50 pixel natural images of landscapes, faces, and animals, as shown in Fig.~\ref{fig:summaryfig}, which is used as our test data, on which the predictions are also applied. 

\begin{figure}[b]
\includegraphics[width=80mm]{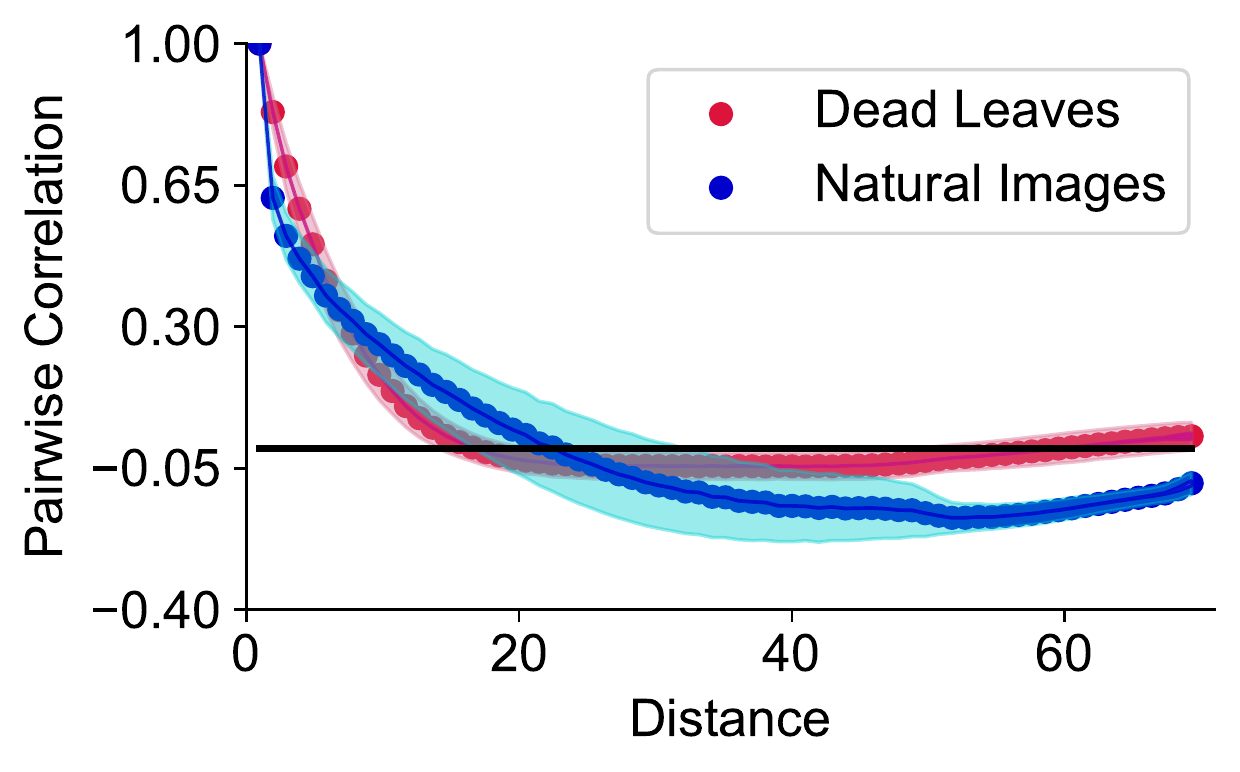}
\caption{\label{fig:correlationplot} \textbf{Pixel-pixel pairwise correlation $c_{{\bf xx}'}$ vs.~pixel-to-pixel distance $d$.} The distances were grouped in 71 evenly spaced bins, since the range of distances in a 50x50 image are between 0 and 70.7. For each bin, we plot the correlation averaged over all pairs of pixels falling into the bin (dots) and the standard deviation (colored bands). Blue -- natural images; red --  dead leaves synthetic data.}
\end{figure}

We consider the correlation between pixels at positions ${\bf x}$ and ${\bf x}'$ in our dataset,
\begin{align}
    c_{{\bf xx}'}&=\frac{1}{T}\sum_{i=1}^T I_i({\bf x})I_i({\bf x}')-\bar{I}({\bf x})\bar{I}({\bf x}'),\\
    \bar{I}({\bf x})& =\frac{1}{T}\sum_{i=1}^TI_i({\bf x}),
\end{align}
where $T$ is the number of images in the dataset, and $\bar{I}({\bf x})$ is the mean value of pixel intensity at position ${\bf x}$ over all images. We expect a strong dependence between the geometric distance of pixels $d^2=(x-x')^2+(y-y')^2$ and $c_{{\bf xx}'}$, with pixels closer to each other being more likely to have the same color. We explore this dependence for both the synthetic and the natural data in Fig.~\ref{fig:correlationplot}. The general structure of $c_{{\bf xx}'}(d)$ curves is similar for both datasets, with a rapid decay to zero indicating a statistical association between the strength of the correlation and the pixel-to-pixel distance. However, there is a distribution of correlations at the same distance $d$ due to the large number of pixel pairs at each distance in the same image. The standard deviation of the correlation over pairs of pixels the same distance apart (denoted as color bands in Fig.~\ref{fig:correlationplot}) measures the inhomogeneity of the images, and it is much larger for the natural data set.

The wide correlation bands in Fig.~\ref{fig:correlationplot} indicate that the relationship between correlation and distance is not deterministic, even at very large data set sizes. This raises questions if it would be possible to use correlations to reconstruct the relative geometric position of pixels. To test this, we destroy the local structure by choosing a random permutation $\pi$, which reshuffles pixel positions, ${\bf x}^*=\pi({\bf x})$. We plot original images $I({\bf x})$ and their shuffled versions  $I({\bf x}^*)$ in Fig.~\ref{fig:summaryfig}.

{\em Reconstructing local relations.} In order to reconstruct the local relations among the shuffled pixels, we employed t-SNE, an algorithm designed to find an optimal embedding of the pixel coordinates in a metric space \cite{vanderMaaten2008}. Other low-dimensional embedding methods could have been used \cite{tenenbaum2000, belkin2003, McInnes2018}, and we view our choice as a demonstration that the reconstruction of the local structure is possible with some methods, and not an endorsement of a specific method, t-SNE in this case.

t-SNE takes as inputs the matrix of Euclidean distances between the variables (pixels in our case) and the desired dimensionality of the embedding. Its output is a set of coordinates for each variable in the constructed embedding.  t-SNE constructs the embedding by minimizing the overall distortion between the geometric arrangement of variables evaluated in the original and in the embedding space, as captured by the Kullback-Leibler divergence, or $D_{\rm KL}$. The distortion is non-uniformly penalized, so that preserving distances between nearby variables is prioritized over those between faraway variables. It is not guaranteed to find a globally optimal embedding, and its solution are controlled by the initial condition and by two additional input parameters: the {\em perplexity} and the {\em early exaggeration} (EE). The former determines the effective number of ``neighbors'' (variables that are allowed to affect the embedding coordinates of a given variable), while the latter controls the global clumpiness of the embedding. We specifically chose t-SNE over other methods designed to produce geometric embeddings of data because strong correlations (small distances) in our data set are well-sampled, whereas small correlations (and hence large distances) suffer from large statistical noise. Thus, preserving them in the embedding is not crucial (cf.~Fig.~\ref{fig:correlationplot}).

To use t-SNE for embedding the shuffled pixels, we first transform their correlation matrix (\textbf{C}) into a distance matrix (\textbf{D}) by \textbf{D} = $ \text{exp} \, (-\textbf{C})$. We tried other transformations, which gave comparable results. Passing the distance matrix $\textbf{D}$ to t-SNE, we get the embedding coordinates (the reconstruction) in the space of the requested dimensionality for each pixel ${\bf x}_{\rm reconst}=f({\bf x}^*)$ in the original data set. If the embedding is in 2d, in addition to the KL divergence, we can evaluate the quality of the reconstruction by directly comparing the original and the reconstructed image coordinates. To avoid issues related to global rotations and the mirror symmetry, which t-SNE cannot recover, we perform the comparison by calculating the correlation coefficient between distances of every pair of pixels in the original image on the one hand, $d({\bf x},{\bf x}')$, and the distances between their coordinates in the reconstructed embedding on the other, $d({\bf x}_{\rm reconst},{\bf x}'_{\rm reconst})=d(f(\pi({\bf x})),f(\pi({\bf x}')))$.

\begin{figure}
\includegraphics[width = 80mm]{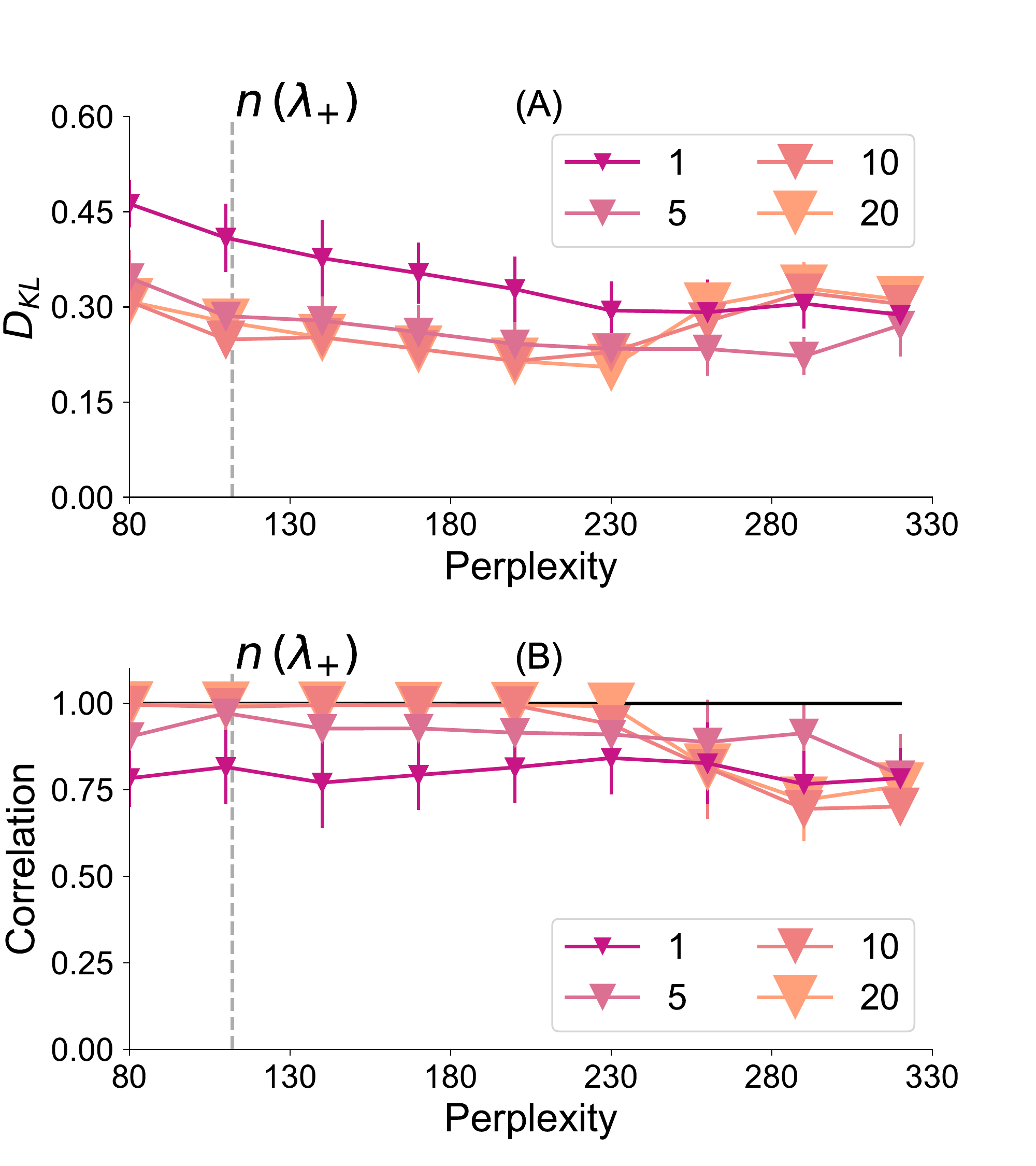}\\
\caption{\label{fig:perplexityspace}
\textbf{DKL and correlation between reconstructed and true distances.} (A) The quality of the optimal 2D embedding found by t-SNE as a function of the early exaggeration (EE) and the perplexity, averaged over 50 runs, with error bars indicating sample standard deviations. The number of eigenvalues above the  statistical significance threshold,  $n(\lambda_+)$, is shown. Different colors/marker sizes indicate different EE values. (B) The average correlation between the pixel-to-pixel distances in the nonpermuted data and t-SNE-embedded permuted data. Plotting notations same as in (A). Dead leaves (training) data is used in both panels. }
\end{figure}

{\em Results.}
To select the appropriate perplexity, we note that larger perplexity values increase the number of neighbors that inform the algorithm about a pixel's position, and hence reduce statistical fluctuations that can come from having a small number of neighbors. However, in our data, the pairwise correlation bands start to dip below zero at a radius of $r\approx 12$ pixels, cf.~Fig.~\ref{fig:correlationplot}. Thus, there are at most about $n_{\rm c}=\pi r^2\approx 450$ pixels that are positively correlated with a pixel far away from the boundary and can be considered its useful neighbors for reconstructing the pixel's position. Pixels at the boundary have even fewer neighbors. Further,  most of $n_{\rm c}$ positive correlations are low, making them uninformative, and potentially even detrimental. Therefore, we expect a good reconstruction of local arrangements with perplexity of only a fraction of $n_{\rm c}$, and a worsening  quality if more neighbors are included. To further develop this intuition, we notice that $T$ images made from $N=T/q<T$ independent pixels will result in nonzero correlations purely due to statistical fluctuations when $q\sim 1$. The celebrated Marchenko-Pastur bound \cite{potters2020first} suggests that we should not trust eigenvalues of covariance matrices smaller than 
\begin{align}
    \lambda_{+} = \sigma^{2} ( 1+ \sqrt{q})^{2},
\end{align}
where $\sigma^2$ is the variance of an individual pixel. On the other hand, correlations stronger that $\lambda_+$ are probably reliably known. In other words, if there are $n(\lambda_+)$ eigenvalues in the pixel-pixel correlation matrix \textbf{C} that are above the $\lambda_+$ cutoff, then $n(\lambda_+)$ linear combinations of pixels can be used reliably to embed a given pixel. Thus  $n(\lambda_+)$ provides the lower estimate on the optimal perplexity range, while $n_{\rm c}$ is the upper bound. As Fig.~\ref{fig:perplexityspace} shows, empirically, the optimal perplexity is closer to $n(\lambda_+)$ than to $n_{\rm c}$, and we will use the optimal perplexity  $p_{\rm opt}=n(\lambda_+)$ in all analyses, unless otherwise specified.  

We know of no heuristics to set a good EE value from the first principles, as a robust theory of hyperparameter selection in t-SNE based on a given data set does not exist. Rather, hyperparameter selection tailored to data sets previously involved trying combinations of hyperparameters (including learning rate and steps) within large ranges \cite{Gove2022, Belkina2019}. Thus, we explore a range of EE values in the analyses below.

Figure \ref{fig:perplexityspace}(A) shows $D_{\rm KL}$ for the quality of the embedding in 2D. For all but the smallest EE values, $D_{\rm KL}$ decreases until the perplexity reaches $\lambda_+$, then stays relatively flat until it starts increasing  well below the naive estimate of 450. Similarly, the correlation between the original distances between pairs of pixels in the non-permuted image and the distances between the same pairs in the embedding space in Fig.~\ref{fig:perplexityspace}(B) shows an improving correlation up to the $\lambda_+$ bound---to nearly perfect values!---and then a drop off soon after, when too many noisy pixel pairs are used to estimate the local structure.  Crucially, these data suggest that small EE is detrimental, but there is little difference in its value past $\sim 10$ in the 2D embedding. 

\begin{figure}
\includegraphics[width=75mm]{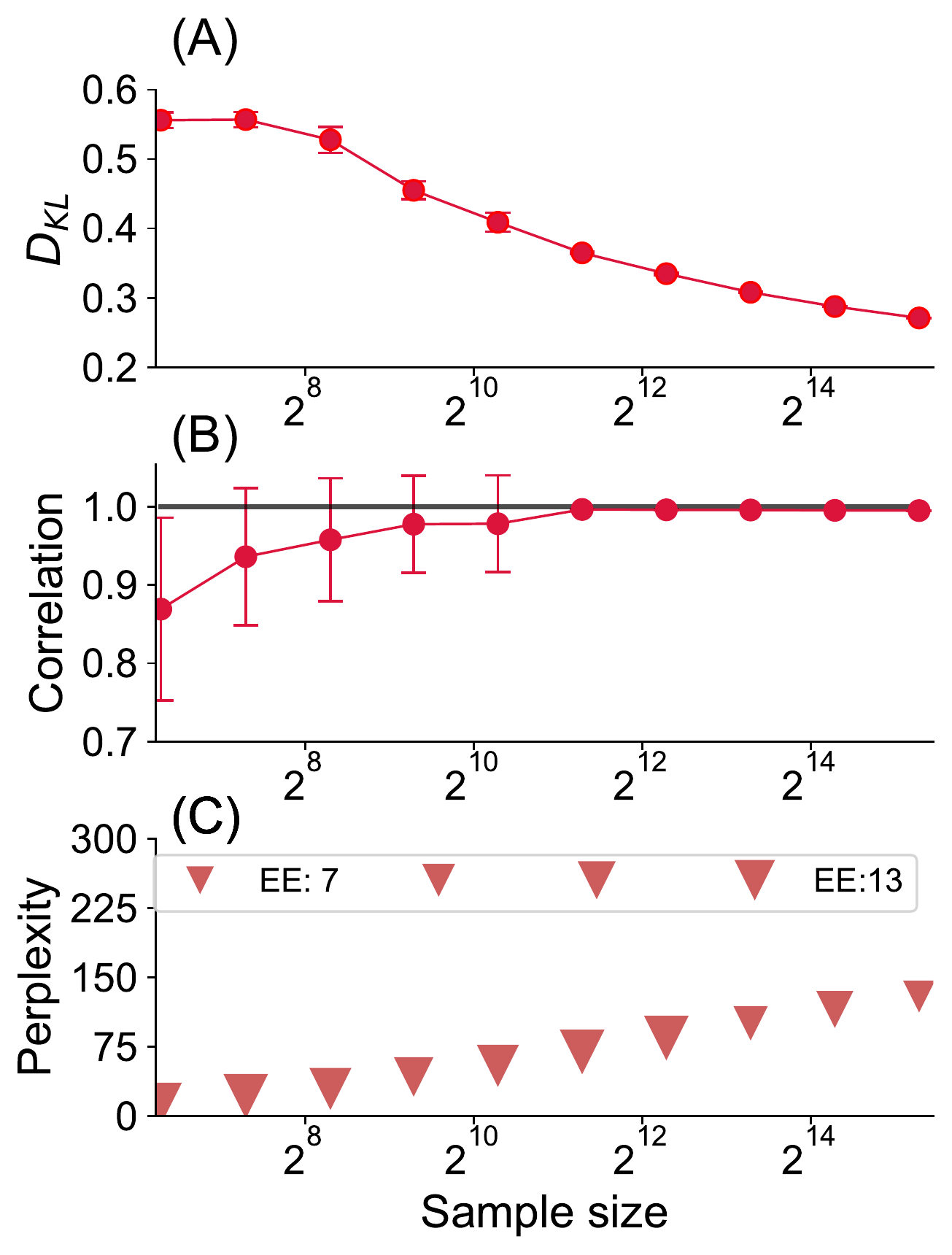}
\caption{\label{fig:datasetvsdivergence} \textbf{Effect of the data set size on the detection of local relations.} (A)  The minimum $D_{\rm KL}$ obtained by t-SNE as a function of the sample size $T$. (B) The correlation the true distances and t-SNE reconstructed distances. Both panels show averages over ten realizations, and error bars too small to be seen in panel (A)) are the standard deviations over the realizations. (C) The optimal perplexity $p_{\rm opt}=n(\lambda_+)$ and the optimal EE value (denoted by the marker size) that achieved the lowest $D_{\rm KL}$ at $p_{\rm opt}$. The optimal EE barely changes over a 1000-fold change in $T$. }
\end{figure}

The weak dependence of neighborhood reconstruction on perplexity suggests that there is enough information about the local structure in just a handful of pairwise distances, so that one can achieve a very good reconstruction even for much smaller sample sizes. Indeed, smaller data sets will have a higher noise in weak correlations between far away pixels. However, a few strong correlations among nearby pixels will still have a small relative error, and this should be sufficient for the reconstruction if a correct  perplexity is chosen. We verify this in Fig.~\ref{fig:datasetvsdivergence}, where we plot the  reconstruction quality as a function of the sample size, $T$. The quality decreases gracefully as $T$ decreases. The average reconstruction correlation reaches $>0.9$ at just $T=157$ images with $n<30$ neighbors used for the reconstruction. In comparison, there are $2500$ pixels and $>3\cdot 10^6$ pairwise distances in the images, so that extremely undersampled data sets are still sufficient for establishing locality.

\begin{figure}
\includegraphics[width = 80mm]{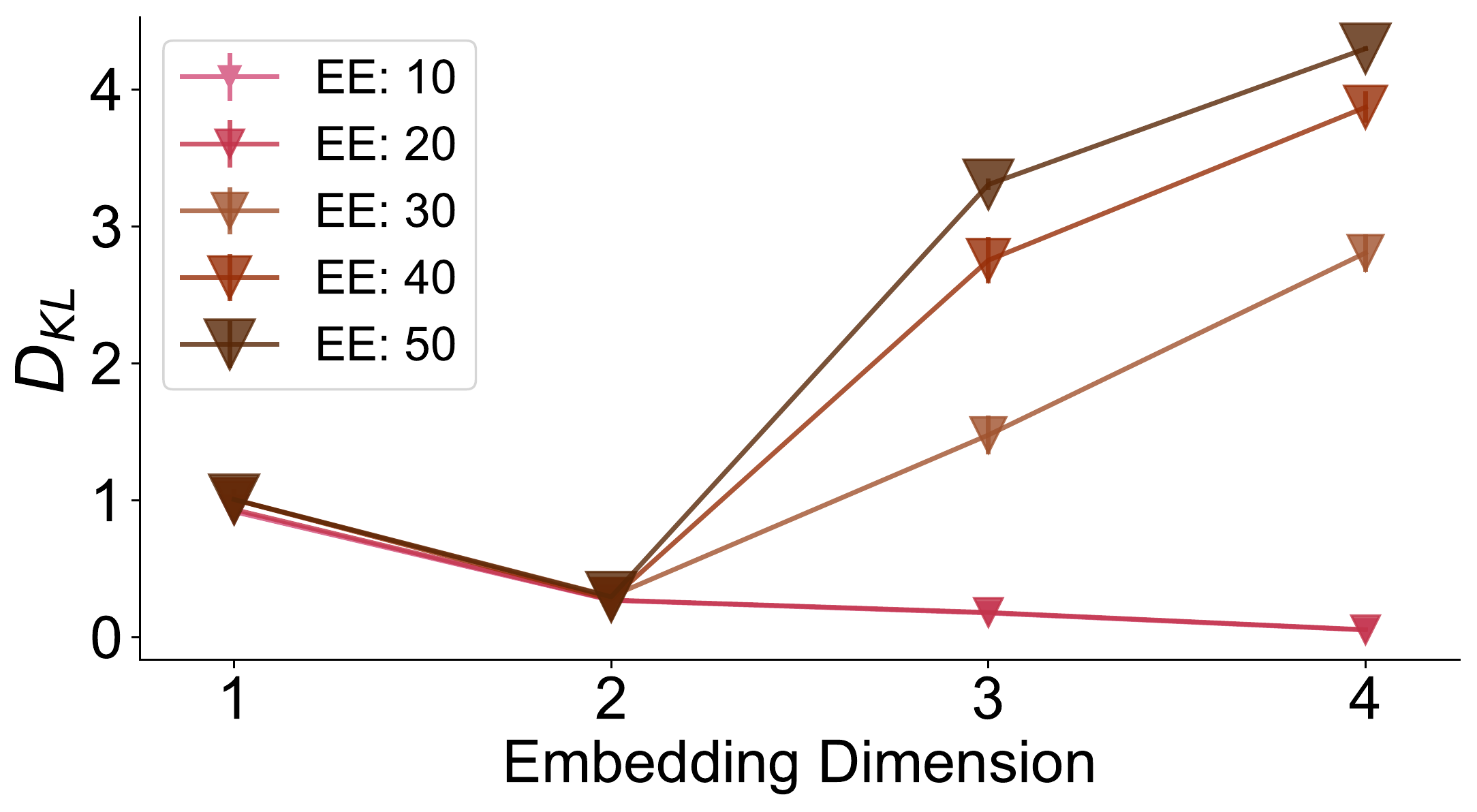}\\
\caption{\label{fig:divergenceplots} \textbf{Embedding quality in different dimensions.} $D_{\rm KL}$ for the optimal embedding is plotted for dimensions of 1 to 4 (averaged over ten realizations, with error bars----too small to see---representing the standard deviation of the set). For our dataset of 2-d images, $D_{\rm KL}$ drops between 1-d and 2-d uniformly for all EEs, and then it becomes strongly susceptible to EE, indicating that the intrinsic dimensionality of the data is 2.  }
\end{figure}

In real data, the correct embedding dimension is often unknown, undefined, or varies across the data set. When possible, it should be inferred from the data directly.  Generically, the larger the embedding dimension, the easier it is for t-SNE to produce embeddings with a smaller $D_{\rm KL}$  simply because it is easier to satisfy distance relations implied by ${\mathbf D}$. This results in the drop in $D_{\rm KL}$ between the embedding dimensions of 1 and 2 in Fig.~\ref{fig:divergenceplots}. However, this freedom comes at a cost that, in higher dimensions, there are many embedding configurations that preserve some relative distances, but distort the global geometry of the image. This makes the reconstruction  susceptible to EE, which controls the clustering of the embedding, increasing the initial distance between clusters while clumping the points within the same cluster. As shown in Fig.~\ref{fig:divergenceplots}, at high EE and in high dimensions, t-SNE fails to recover from extremely clumpy initializations, resulting in large  $D_{\rm KL}$. Only at or below the true dimensionality of 2 is the algorithm insensitive to EE. This suggests a simple approach to detecting the intrinsic dimensionality of data: look for the largest embedding dimension, at which the reconstruction $D_{\rm KL}$ is the lowest uniformly, independently of the EE. We emphasize this finding: with no assumptions about  the structure of the data, t-SNE can recover its intrinsic dimensionality. This bodes well for applications to more interesting data sets. However, Cauchy distributions, which allow t-SNE to preserve local structures while breaking global relations, do not work well in higher dimensions \cite{vanderMaaten2008}. Thus applicability of this simple approach to large-dimensional data requires additional investigation.

Finally, to illustrate the weak sensitivity of the recovery of the local relations to the nature of the training set, we visualize the reconstructed images. For this, we place a grey or a black dot, corresponding to the color of the shuffled pixel in the original image, at the optimal 2-d t-SNE reconstructed coordinates ${\bf x}_{\rm recon}$. We then rescale the reconstructed image to be 50 x 50 pixels in size (matching the original images), and rotate them to better align to the original images (notice that the rescaling and the rotation are the same for {\em all} images). This is needed because t-SNE focuses on local geometry only and is unaware of possible global transformations to the data. We show the results for a few select training and test images in the right column of Fig.~\ref{fig:summaryfig}. The quality of the reconstruction for both data sets is visually exceptional, even though the correlation structure is different in both data sets, cf.~Fig.~\ref{fig:correlationplot}. 

{\em Discussion.}
Building probabilistic models $P(I(\bf x))$ of large dimensional biological systems of realistic sizes is a combinatorial challenge, especially as $N > T$.  Regularizing the learning problem requires knowing local relations and hence  which variables are allowed to interact. Such relations may or may not correspond to neighborhoods in the real space, but some notion of effective locality is crucial for model building. Here we showed that, at least in the context of a 2-d images, pairwise correlations among observed variables alone are sufficient to recover the locality structure and the dimensionality of the data set using an off-the-shelf visualization algorithm, t-SNE (though we emphasize again that we view t-SNE not as a crucial choice of our approach, but as just one of the methods that we could have used). We provided semi-quantitative heuristics based on well-known calculations from the random matrix theory to set parameters of t-SNE to determine the optimal embedding dimension and the optimal reconstruction for this dimension.  The resulting inference is so good that, even in the deeply undersampled regime, the permuted images can be fully reconstructed, with a very weak sensitivity to the details of the training set. 

The key insight from this analysis is that, even when the sample size is insufficient to estimate all correlations in data, a few strong correlations can still be estimated well. The number of such correlations is then enough to predict the position of a pixel relative to its ``neighbors'', even if relations to the more distant pixels are unknown. One can then reconstruct the full structure of the data set by traversing the neighborhoods of the pixels individually, --- which is what t-SNE (and other related low-dimensional visualization algorithms) does implicitly.

Real-world data are usually more complex than 2-d images considered here. Dimensionality of such data often depends on the point in the state space and is generally hard to define (see, e.g., Ref.~\cite{berman} for discussion of this for animal behavior data). Common generative models of real data often are based on low-dimensional or sparse {\em latent} (rather than visible) structures (see, e.g., Ref.~\cite{vyas2020} for a relevant discussion in the context of neurobiology). Such latent models also can generate data of varying or undefined global dimensionality in the visible space. Making generalization from a relatively simple problem of 2-d images to such complex data is hard. However, we expect that, while direct application of t-SNE or other algorithms that assume a small fixed dimensionality to more complex data may be questionable, our main realization that determining {\em local} structures only requires a few strong correlations irrespective of the {\em global} properties of the data is likely to hold for them as well. Thus, we expect that determining local relations and then constraining statistical models to only include interactions within them will decrease the number of parameters that must be inferred to build a model with $N$ variables from  $O(2^N)$ to $O(N)$, making such models experimentally tractable. Applications to real data in diverse domains can be plentiful, as possibilities include identifying functional 3D protein structure from co-evolution based amino acids \cite{Marks2011}, chromosomal structure based on base-pair contacts \cite{Lieberman-Aiden2009}, neural activity from correlated firing patterns \cite{Savin2017}, or ``favored'' interactions in genome space \cite{Mora2010}. All of these biological processes are similar in that the number of  possible combinatorial interactions greatly exceeds possible sizes of experimental datasets. Unknown physical constraints create an effective locality, which is distinct from the geometric locality (e.~g., allostery allows for long-range interactions in proteins), and it is precisely this effective locality---{\em a priori} unknown---that offers a possibility for model building. Understanding for which of these domains, if any, our approach works or not and why is likely to be a fruitful direction for future research.

Success of modern machine learning in image analysis has largely been driven by convolutional neural networks, which decrease the dimensionality of the learned statistical model by imposing locality and translational invariance on images \cite{rawat2017deep}. Our analysis suggests that the local structure can be inferred from just a handful of images. Thus, there should exist algorithms for training general fully connected neural networks so that, compared to the convolutional networks, they are only minimally handicapped by not knowing the structure {\em a priori}.

We point out an intriguing speculative connection to the transformer neural architecture \cite{transformer} and, more generally, other machine learning models with the so called {\em attention} mechanism \cite{attention}. Transformers are behind the recent success of Large Language Models, such as GPT-3 \cite{GPT3}. They build on an established idea that the identity of a word that should occur at a certain place in text can be determined by the context---that is, by the conditional distribution of a word on its surroundings \cite{pereira1994distributional}. However, contexts, specified by identities of words and their relative position in text, must be very large to be practically useful. Hence they are severely undersampled even for exceptionally large training sets. Transformers solve this problem by using the attention mechanism to only focus on the parts of the context, whose correlations with the target are well-sampled. A potential future research direction involves exploring if such attention is similar to how we were able to detect an attribute (position) of a pixel by only focusing on highly correlated pixels and ignoring noisy, undersampled relations. If a rigorous connection can be identified, one may be able to use random matrix theory-based estimates, similar to our current analysis, to understand the data set sizes needed for attention-based models to work.

\begin{acknowledgements}
We are grateful to Sean Ridout for insightful comments about the manuscript. IN was supported, in part, by the Simons Investigator Award. MR  was supported, in part, by Emory University internal funds. 
\end{acknowledgements}

\bibliography{apssamp}

\begin{thebibliography}{30}%
\makeatletter
\providecommand \@ifxundefined [1]{%
 \@ifx{#1\undefined}
}%
\providecommand \@ifnum [1]{%
 \ifnum #1\expandafter \@firstoftwo
 \else \expandafter \@secondoftwo
 \fi
}%
\providecommand \@ifx [1]{%
 \ifx #1\expandafter \@firstoftwo
 \else \expandafter \@secondoftwo
 \fi
}%
\providecommand \natexlab [1]{#1}%
\providecommand \enquote  [1]{``#1''}%
\providecommand \bibnamefont  [1]{#1}%
\providecommand \bibfnamefont [1]{#1}%
\providecommand \citenamefont [1]{#1}%
\providecommand \href@noop [0]{\@secondoftwo}%
\providecommand \href [0]{\begingroup \@sanitize@url \@href}%
\providecommand \@href[1]{\@@startlink{#1}\@@href}%
\providecommand \@@href[1]{\endgroup#1\@@endlink}%
\providecommand \@sanitize@url [0]{\catcode `\\12\catcode `\$12\catcode
  `\&12\catcode `\#12\catcode `\^12\catcode `\_12\catcode `\%12\relax}%
\providecommand \@@startlink[1]{}%
\providecommand \@@endlink[0]{}%
\providecommand \url  [0]{\begingroup\@sanitize@url \@url }%
\providecommand \@url [1]{\endgroup\@href {#1}{\urlprefix }}%
\providecommand \urlprefix  [0]{URL }%
\providecommand \Eprint [0]{\href }%
\providecommand \doibase [0]{https://doi.org/}%
\providecommand \selectlanguage [0]{\@gobble}%
\providecommand \bibinfo  [0]{\@secondoftwo}%
\providecommand \bibfield  [0]{\@secondoftwo}%
\providecommand \translation [1]{[#1]}%
\providecommand \BibitemOpen [0]{}%
\providecommand \bibitemStop [0]{}%
\providecommand \bibitemNoStop [0]{.\EOS\space}%
\providecommand \EOS [0]{\spacefactor3000\relax}%
\providecommand \BibitemShut  [1]{\csname bibitem#1\endcsname}%
\let\auto@bib@innerbib\@empty
\bibitem [{\citenamefont {Carleo}\ \emph {et~al.}(2019)\citenamefont {Carleo},
  \citenamefont {Cirac}, \citenamefont {Cranmer}, \citenamefont {Daudet},
  \citenamefont {Schuld}, \citenamefont {Tishby}, \citenamefont
  {Vogt-Maranto},\ and\ \citenamefont {Zdeborov{\'a}}}]{carleo2019machine}%
  \BibitemOpen
  \bibfield  {author} {\bibinfo {author} {\bibfnamefont {G.}~\bibnamefont
  {Carleo}}, \bibinfo {author} {\bibfnamefont {I.}~\bibnamefont {Cirac}},
  \bibinfo {author} {\bibfnamefont {K.}~\bibnamefont {Cranmer}}, \bibinfo
  {author} {\bibfnamefont {L.}~\bibnamefont {Daudet}}, \bibinfo {author}
  {\bibfnamefont {M.}~\bibnamefont {Schuld}}, \bibinfo {author} {\bibfnamefont
  {N.}~\bibnamefont {Tishby}}, \bibinfo {author} {\bibfnamefont
  {L.}~\bibnamefont {Vogt-Maranto}},\ and\ \bibinfo {author} {\bibfnamefont
  {L.}~\bibnamefont {Zdeborov{\'a}}},\ }\bibfield  {title} {\bibinfo {title}
  {Machine learning and the physical sciences},\ }\href@noop {} {\bibfield
  {journal} {\bibinfo  {journal} {Rev.\ Mod.\ Phys.}\ }\textbf {\bibinfo
  {volume} {91}},\ \bibinfo {pages} {045002} (\bibinfo {year}
  {2019})}\BibitemShut {NoStop}%
\bibitem [{\citenamefont {Savin}\ and\ \citenamefont {Tka{\v
  c}ik}(2017)}]{Savin2017}%
  \BibitemOpen
  \bibfield  {author} {\bibinfo {author} {\bibfnamefont {C.}~\bibnamefont
  {Savin}}\ and\ \bibinfo {author} {\bibfnamefont {G.}~\bibnamefont {Tka{\v
  c}ik}},\ }\bibfield  {title} {\bibinfo {title} {Maximum entropy models as a
  tool for building precise neural controls},\ }\href@noop {} {\bibfield
  {journal} {\bibinfo  {journal} {Curr.\ Opin.\ Neurobiol.}\ }\textbf {\bibinfo
  {volume} {46}},\ \bibinfo {pages} {120} (\bibinfo {year} {2017})}\BibitemShut
  {NoStop}%
\bibitem [{\citenamefont {Marks}\ \emph {et~al.}(2011)\citenamefont {Marks},
  \citenamefont {Colwell}, \citenamefont {Hopf}, \citenamefont {Pagnani},
  \citenamefont {Zecchina},\ and\ \citenamefont {Sander}}]{Marks2011}%
  \BibitemOpen
  \bibfield  {author} {\bibinfo {author} {\bibfnamefont {D.~S.}\ \bibnamefont
  {Marks}}, \bibinfo {author} {\bibfnamefont {L.~J.}\ \bibnamefont {Colwell}},
  \bibinfo {author} {\bibfnamefont {T.~A.}\ \bibnamefont {Hopf}}, \bibinfo
  {author} {\bibfnamefont {A.}~\bibnamefont {Pagnani}}, \bibinfo {author}
  {\bibfnamefont {R.}~\bibnamefont {Zecchina}},\ and\ \bibinfo {author}
  {\bibfnamefont {C.}~\bibnamefont {Sander}},\ }\bibfield  {title} {\bibinfo
  {title} {Protein 3d structure computed from evolutionary sequence
  variation},\ }\href@noop {} {\bibfield  {journal} {\bibinfo  {journal} {PLoS
  One}\ }\textbf {\bibinfo {volume} {6}},\ \bibinfo {pages} {35} (\bibinfo
  {year} {2011})}\BibitemShut {NoStop}%
\bibitem [{\citenamefont {Donoho}(2006)}]{Donoho2006}%
  \BibitemOpen
  \bibfield  {author} {\bibinfo {author} {\bibfnamefont {D.~L.}\ \bibnamefont
  {Donoho}},\ }\bibfield  {title} {\bibinfo {title} {Compressed sensing},\
  }\href@noop {} {\bibfield  {journal} {\bibinfo  {journal} {IEEE Trans.\ Inf.\
  Thy.}\ }\textbf {\bibinfo {volume} {52}},\ \bibinfo {pages} {1289} (\bibinfo
  {year} {2006})}\BibitemShut {NoStop}%
\bibitem [{\citenamefont {Bailly-Bechet}\ \emph {et~al.}(2010)\citenamefont
  {Bailly-Bechet}, \citenamefont {Braunstein}, \citenamefont {Pagnani},
  \citenamefont {Weigt},\ and\ \citenamefont {Zecchina}}]{bailly-bechet2010}%
  \BibitemOpen
  \bibfield  {author} {\bibinfo {author} {\bibfnamefont {M.}~\bibnamefont
  {Bailly-Bechet}}, \bibinfo {author} {\bibfnamefont {A.}~\bibnamefont
  {Braunstein}}, \bibinfo {author} {\bibfnamefont {A.}~\bibnamefont {Pagnani}},
  \bibinfo {author} {\bibfnamefont {M.}~\bibnamefont {Weigt}},\ and\ \bibinfo
  {author} {\bibfnamefont {R.}~\bibnamefont {Zecchina}},\ }\bibfield  {title}
  {\bibinfo {title} {Inference of sparse combinatorial-control networks from
  gene-expression data: a message passing approach},\ }\href@noop {} {\bibfield
   {journal} {\bibinfo  {journal} {BMC Bioinf.}\ }\textbf {\bibinfo {volume}
  {11}} (\bibinfo {year} {2010})}\BibitemShut {NoStop}%
\bibitem [{\citenamefont {Ganguli}\ and\ \citenamefont
  {Sompolinsky}(2012)}]{Ganguli2012}%
  \BibitemOpen
  \bibfield  {author} {\bibinfo {author} {\bibfnamefont {S.}~\bibnamefont
  {Ganguli}}\ and\ \bibinfo {author} {\bibfnamefont {H.}~\bibnamefont
  {Sompolinsky}},\ }\bibfield  {title} {\bibinfo {title} {Compressed sensing,
  sparsity, and dimensionality in neuronal information processing and data
  analysis},\ }\href@noop {} {\bibfield  {journal} {\bibinfo  {journal} {Ann.\
  Rev.\ Neurosci.}\ }\textbf {\bibinfo {volume} {35}},\ \bibinfo {pages} {485}
  (\bibinfo {year} {2012})}\BibitemShut {NoStop}%
\bibitem [{\citenamefont {Bulso}\ \emph {et~al.}(2016)\citenamefont {Bulso},
  \citenamefont {Marsili},\ and\ \citenamefont {Roudi}}]{Bulso2016}%
  \BibitemOpen
  \bibfield  {author} {\bibinfo {author} {\bibfnamefont {N.}~\bibnamefont
  {Bulso}}, \bibinfo {author} {\bibfnamefont {M.}~\bibnamefont {Marsili}},\
  and\ \bibinfo {author} {\bibfnamefont {Y.}~\bibnamefont {Roudi}},\ }\bibfield
   {title} {\bibinfo {title} {Sparse model selection in the highly
  under-sampled regime},\ }\href@noop {} {\bibfield  {journal} {\bibinfo
  {journal} {Journal of Statistical Mechanics: Theory and Experiment}\ }\textbf
  {\bibinfo {volume} {2016}} (\bibinfo {year} {2016})}\BibitemShut {NoStop}%
\bibitem [{\citenamefont {Girvan}\ and\ \citenamefont
  {Newman}(2002)}]{girvan2001}%
  \BibitemOpen
  \bibfield  {author} {\bibinfo {author} {\bibfnamefont {M.}~\bibnamefont
  {Girvan}}\ and\ \bibinfo {author} {\bibfnamefont {M.~E.~J.}\ \bibnamefont
  {Newman}},\ }\bibfield  {title} {\bibinfo {title} {Community structure in
  social and biological networks},\ }\href@noop {} {\bibfield  {journal}
  {\bibinfo  {journal} {Proc.\ Natl.\ Acad.\ Sci.\ (USA)}\ }\textbf {\bibinfo
  {volume} {99}},\ \bibinfo {pages} {7821} (\bibinfo {year}
  {2002})}\BibitemShut {NoStop}%
\bibitem [{\citenamefont {Clauset}(2005)}]{Clauset2005}%
  \BibitemOpen
  \bibfield  {author} {\bibinfo {author} {\bibfnamefont {A.}~\bibnamefont
  {Clauset}},\ }\bibfield  {title} {\bibinfo {title} {Finding local community
  structure in networks},\ }\href@noop {} {\bibfield  {journal} {\bibinfo
  {journal} {Phys.\ Rev.\ E}\ }\textbf {\bibinfo {volume} {72}} (\bibinfo
  {year} {2005})}\BibitemShut {NoStop}%
\bibitem [{\citenamefont {Ravasz}\ and\ \citenamefont
  {Barab{\'a}si}(2003)}]{Ravasz2003}%
  \BibitemOpen
  \bibfield  {author} {\bibinfo {author} {\bibfnamefont {E.}~\bibnamefont
  {Ravasz}}\ and\ \bibinfo {author} {\bibfnamefont {A.}~\bibnamefont
  {Barab{\'a}si}},\ }\bibfield  {title} {\bibinfo {title} {Hierarchical
  organization in complex networks},\ }\href@noop {} {\bibfield  {journal}
  {\bibinfo  {journal} {Phys.\ Rev.\ E}\ }\textbf {\bibinfo {volume} {67}},\
  \bibinfo {pages} {7} (\bibinfo {year} {2003})}\BibitemShut {NoStop}%
\bibitem [{\citenamefont {Potters}\ and\ \citenamefont
  {Bouchaud}(2020)}]{potters2020first}%
  \BibitemOpen
  \bibfield  {author} {\bibinfo {author} {\bibfnamefont {M.}~\bibnamefont
  {Potters}}\ and\ \bibinfo {author} {\bibfnamefont {J.-P.}\ \bibnamefont
  {Bouchaud}},\ }\href@noop {} {\emph {\bibinfo {title} {A First Course in
  Random Matrix Theory: For Physicists, Engineers and Data Scientists}}}\
  (\bibinfo  {publisher} {Cambridge UP},\ \bibinfo {year} {2020})\BibitemShut
  {NoStop}%
\bibitem [{\citenamefont {Ruderman}\ and\ \citenamefont
  {Bialek}(1994)}]{ruderman1994statistics}%
  \BibitemOpen
  \bibfield  {author} {\bibinfo {author} {\bibfnamefont {D.}~\bibnamefont
  {Ruderman}}\ and\ \bibinfo {author} {\bibfnamefont {W.}~\bibnamefont
  {Bialek}},\ }\bibfield  {title} {\bibinfo {title} {Statistics of natural
  images: Scaling in the woods},\ }\href@noop {} {\bibfield  {journal}
  {\bibinfo  {journal} {Phys.\ Rev.\ Lett.}\ }\textbf {\bibinfo {volume}
  {73}},\ \bibinfo {pages} {814} (\bibinfo {year} {1994})}\BibitemShut
  {NoStop}%
\bibitem [{\citenamefont {Matheron}(1974)}]{Matheron1974}%
  \BibitemOpen
  \bibfield  {author} {\bibinfo {author} {\bibfnamefont {G.}~\bibnamefont
  {Matheron}},\ }\href@noop {} {\emph {\bibinfo {title} {Random Sets and
  Integral Geometry}}}\ (\bibinfo  {publisher} {John Wiley and Sons},\ \bibinfo
  {year} {1974})\BibitemShut {NoStop}%
\bibitem [{\citenamefont {Lee}\ \emph {et~al.}(2001)\citenamefont {Lee},
  \citenamefont {Mumford},\ and\ \citenamefont {Huang}}]{Lee2001}%
  \BibitemOpen
  \bibfield  {author} {\bibinfo {author} {\bibfnamefont {A.~B.}\ \bibnamefont
  {Lee}}, \bibinfo {author} {\bibfnamefont {D.}~\bibnamefont {Mumford}},\ and\
  \bibinfo {author} {\bibfnamefont {J.}~\bibnamefont {Huang}},\ }\bibfield
  {title} {\bibinfo {title} {Occlusion models for natural images: A statistical
  study of a scale-invariant dead leaves model},\ }\href@noop {} {\bibfield
  {journal} {\bibinfo  {journal} {Int.\ J.\ Comput.\ Vis.}\ }\textbf {\bibinfo
  {volume} {41}},\ \bibinfo {pages} {35} (\bibinfo {year} {2001})}\BibitemShut
  {NoStop}%
\bibitem [{\citenamefont {Pitkow}(2010)}]{Pitkow2010}%
  \BibitemOpen
  \bibfield  {author} {\bibinfo {author} {\bibfnamefont {X.}~\bibnamefont
  {Pitkow}},\ }\bibfield  {title} {\bibinfo {title} {Exact feature
  probabilities in images with occlusion},\ }\href@noop {} {\bibfield
  {journal} {\bibinfo  {journal} {J.\ Vis.}\ }\textbf {\bibinfo {volume}
  {10}},\ \bibinfo {pages} {1} (\bibinfo {year} {2010})}\BibitemShut {NoStop}%
\bibitem [{\citenamefont {van~der Maaten}\ and\ \citenamefont
  {Hinton}(2008)}]{vanderMaaten2008}%
  \BibitemOpen
  \bibfield  {author} {\bibinfo {author} {\bibfnamefont {L.}~\bibnamefont
  {van~der Maaten}}\ and\ \bibinfo {author} {\bibfnamefont {G.}~\bibnamefont
  {Hinton}},\ }\bibfield  {title} {\bibinfo {title} {Visualizing data using
  t-sne},\ }\href@noop {} {\bibfield  {journal} {\bibinfo  {journal} {J.\
  Machine Learn.\ Res.}\ }\textbf {\bibinfo {volume} {9}},\ \bibinfo {pages}
  {2579} (\bibinfo {year} {2008})}\BibitemShut {NoStop}%
\bibitem [{\citenamefont {Tenenbaum}\ \emph {et~al.}(2000)\citenamefont
  {Tenenbaum}, \citenamefont {de~Silva},\ and\ \citenamefont
  {Langford}}]{tenenbaum2000}%
  \BibitemOpen
  \bibfield  {author} {\bibinfo {author} {\bibfnamefont {J.}~\bibnamefont
  {Tenenbaum}}, \bibinfo {author} {\bibfnamefont {V.}~\bibnamefont
  {de~Silva}},\ and\ \bibinfo {author} {\bibfnamefont {J.}~\bibnamefont
  {Langford}},\ }\bibfield  {title} {\bibinfo {title} {A global geometric
  framework for nonlinear dimensionality reduction},\ }\href@noop {} {\bibfield
   {journal} {\bibinfo  {journal} {Science}\ }\textbf {\bibinfo {volume}
  {290}},\ \bibinfo {pages} {2319} (\bibinfo {year} {2000})}\BibitemShut
  {NoStop}%
\bibitem [{\citenamefont {Belkin}\ and\ \citenamefont
  {Niyogi}(2003)}]{belkin2003}%
  \BibitemOpen
  \bibfield  {author} {\bibinfo {author} {\bibfnamefont {M.}~\bibnamefont
  {Belkin}}\ and\ \bibinfo {author} {\bibfnamefont {P.}~\bibnamefont
  {Niyogi}},\ }\bibfield  {title} {\bibinfo {title} {Laplacian eigenmaps for
  dimensionality reduction and data representation},\ }\href@noop {} {\bibfield
   {journal} {\bibinfo  {journal} {Neural Computation}\ }\textbf {\bibinfo
  {volume} {15}},\ \bibinfo {pages} {1373–} (\bibinfo {year}
  {2003})}\BibitemShut {NoStop}%
\bibitem [{\citenamefont {McInnes}\ \emph {et~al.}(2018)\citenamefont
  {McInnes}, \citenamefont {Healy}, \citenamefont {Saul},\ and\ \citenamefont
  {Großberger}}]{McInnes2018}%
  \BibitemOpen
  \bibfield  {author} {\bibinfo {author} {\bibfnamefont {L.}~\bibnamefont
  {McInnes}}, \bibinfo {author} {\bibfnamefont {J.}~\bibnamefont {Healy}},
  \bibinfo {author} {\bibfnamefont {N.}~\bibnamefont {Saul}},\ and\ \bibinfo
  {author} {\bibfnamefont {L.}~\bibnamefont {Großberger}},\ }\bibfield
  {title} {\bibinfo {title} {Umap: Uniform manifold approximation and
  projection},\ }\href {https://doi.org/10.21105/joss.00861} {\bibfield
  {journal} {\bibinfo  {journal} {Journal of Open Source Software}\ }\textbf
  {\bibinfo {volume} {3}},\ \bibinfo {pages} {861} (\bibinfo {year}
  {2018})}\BibitemShut {NoStop}%
\bibitem [{\citenamefont {Gove}\ \emph {et~al.}(2022)\citenamefont {Gove},
  \citenamefont {Cadalzo}, \citenamefont {Leiby}, \citenamefont {Singer},\ and\
  \citenamefont {Zaitzeff}}]{Gove2022}%
  \BibitemOpen
  \bibfield  {author} {\bibinfo {author} {\bibfnamefont {R.}~\bibnamefont
  {Gove}}, \bibinfo {author} {\bibfnamefont {L.}~\bibnamefont {Cadalzo}},
  \bibinfo {author} {\bibfnamefont {N.}~\bibnamefont {Leiby}}, \bibinfo
  {author} {\bibfnamefont {J.~M.}\ \bibnamefont {Singer}},\ and\ \bibinfo
  {author} {\bibfnamefont {A.}~\bibnamefont {Zaitzeff}},\ }\bibfield  {title}
  {\bibinfo {title} {New guidance for using t-sne: Alternative defaults,
  hyperparameter selection automation, and comparative evaluation},\
  }\href@noop {} {\bibfield  {journal} {\bibinfo  {journal} {Visual
  Informatics}\ }\textbf {\bibinfo {volume} {6}},\ \bibinfo {pages} {87}
  (\bibinfo {year} {2022})}\BibitemShut {NoStop}%
\bibitem [{\citenamefont {Belkina}\ \emph {et~al.}(2019)\citenamefont
  {Belkina}, \citenamefont {Ciccolella}, \citenamefont {Anno}, \citenamefont
  {Halpert}, \citenamefont {Spidlen},\ and\ \citenamefont
  {Snyder-Cappione}}]{Belkina2019}%
  \BibitemOpen
  \bibfield  {author} {\bibinfo {author} {\bibfnamefont {A.~C.}\ \bibnamefont
  {Belkina}}, \bibinfo {author} {\bibfnamefont {C.~O.}\ \bibnamefont
  {Ciccolella}}, \bibinfo {author} {\bibfnamefont {R.}~\bibnamefont {Anno}},
  \bibinfo {author} {\bibfnamefont {R.}~\bibnamefont {Halpert}}, \bibinfo
  {author} {\bibfnamefont {J.}~\bibnamefont {Spidlen}},\ and\ \bibinfo {author}
  {\bibfnamefont {J.~E.}\ \bibnamefont {Snyder-Cappione}},\ }\bibfield  {title}
  {\bibinfo {title} {Automated optimized parameters for t-distributed
  stochastic neighbor embedding improve visualization and analysis of large
  datasets},\ }\href@noop {} {\bibfield  {journal} {\bibinfo  {journal} {Nature
  Commun.}\ }\textbf {\bibinfo {volume} {10}} (\bibinfo {year}
  {2019})}\BibitemShut {NoStop}%
\bibitem [{\citenamefont {Berman}(2018)}]{berman}%
  \BibitemOpen
  \bibfield  {author} {\bibinfo {author} {\bibfnamefont {G.}~\bibnamefont
  {Berman}},\ }\bibfield  {title} {\bibinfo {title} {Measuring behavior across
  scales},\ }\href@noop {} {\bibfield  {journal} {\bibinfo  {journal} {BMC
  Biol}\ }\textbf {\bibinfo {volume} {16}},\ \bibinfo {pages} {23} (\bibinfo
  {year} {2018})}\BibitemShut {NoStop}%
\bibitem [{\citenamefont {Vyas}\ \emph {et~al.}(2020)\citenamefont {Vyas},
  \citenamefont {Golub}, \citenamefont {Sussillo},\ and\ \citenamefont
  {Shenoy}}]{vyas2020}%
  \BibitemOpen
  \bibfield  {author} {\bibinfo {author} {\bibfnamefont {S.}~\bibnamefont
  {Vyas}}, \bibinfo {author} {\bibfnamefont {M.}~\bibnamefont {Golub}},
  \bibinfo {author} {\bibfnamefont {D.}~\bibnamefont {Sussillo}},\ and\
  \bibinfo {author} {\bibfnamefont {K.}~\bibnamefont {Shenoy}},\ }\bibfield
  {title} {\bibinfo {title} {Computation through neural population dynamics},\
  }\href@noop {} {\bibfield  {journal} {\bibinfo  {journal} {Annu.\ Rev.\
  Neurosci.}\ }\textbf {\bibinfo {volume} {43}},\ \bibinfo {pages} {249}
  (\bibinfo {year} {2020})}\BibitemShut {NoStop}%
\bibitem [{\citenamefont {Lieberman-Aiden}\ \emph {et~al.}(2009)\citenamefont
  {Lieberman-Aiden}, \citenamefont {Van~Berkum}, \citenamefont {Williams},
  \citenamefont {Imakaev}, \citenamefont {Ragoczy}, \citenamefont {Telling},
  \citenamefont {Amit}, \citenamefont {Lajoie}, \citenamefont {Sabo},
  \citenamefont {Dorschner} \emph {et~al.}}]{Lieberman-Aiden2009}%
  \BibitemOpen
  \bibfield  {author} {\bibinfo {author} {\bibfnamefont {E.}~\bibnamefont
  {Lieberman-Aiden}}, \bibinfo {author} {\bibfnamefont {N.~L.}\ \bibnamefont
  {Van~Berkum}}, \bibinfo {author} {\bibfnamefont {L.}~\bibnamefont
  {Williams}}, \bibinfo {author} {\bibfnamefont {M.}~\bibnamefont {Imakaev}},
  \bibinfo {author} {\bibfnamefont {T.}~\bibnamefont {Ragoczy}}, \bibinfo
  {author} {\bibfnamefont {A.}~\bibnamefont {Telling}}, \bibinfo {author}
  {\bibfnamefont {I.}~\bibnamefont {Amit}}, \bibinfo {author} {\bibfnamefont
  {B.~R.}\ \bibnamefont {Lajoie}}, \bibinfo {author} {\bibfnamefont {P.~J.}\
  \bibnamefont {Sabo}}, \bibinfo {author} {\bibfnamefont {M.~O.}\ \bibnamefont
  {Dorschner}}, \emph {et~al.},\ }\bibfield  {title} {\bibinfo {title}
  {Comprehensive mapping of long-range interactions reveals folding principles
  of the human genome},\ }\href@noop {} {\bibfield  {journal} {\bibinfo
  {journal} {Science}\ }\textbf {\bibinfo {volume} {326}},\ \bibinfo {pages}
  {289} (\bibinfo {year} {2009})}\BibitemShut {NoStop}%
\bibitem [{\citenamefont {Mora}\ \emph {et~al.}(2010)\citenamefont {Mora},
  \citenamefont {Walczak}, \citenamefont {Bialek},\ and\ \citenamefont
  {Callan}}]{Mora2010}%
  \BibitemOpen
  \bibfield  {author} {\bibinfo {author} {\bibfnamefont {T.}~\bibnamefont
  {Mora}}, \bibinfo {author} {\bibfnamefont {A.}~\bibnamefont {Walczak}},
  \bibinfo {author} {\bibfnamefont {W.}~\bibnamefont {Bialek}},\ and\ \bibinfo
  {author} {\bibfnamefont {C.}~\bibnamefont {Callan}},\ }\bibfield  {title}
  {\bibinfo {title} {Maximum entropy models for antibody diversity},\
  }\href@noop {} {\bibfield  {journal} {\bibinfo  {journal} {Proc.\ Natl.\
  Acad.\ Sci.\ (USA)}\ }\textbf {\bibinfo {volume} {107}},\ \bibinfo {pages}
  {5405} (\bibinfo {year} {2010})}\BibitemShut {NoStop}%
\bibitem [{\citenamefont {Rawat}\ and\ \citenamefont
  {Wang}(2017)}]{rawat2017deep}%
  \BibitemOpen
  \bibfield  {author} {\bibinfo {author} {\bibfnamefont {W.}~\bibnamefont
  {Rawat}}\ and\ \bibinfo {author} {\bibfnamefont {Z.}~\bibnamefont {Wang}},\
  }\bibfield  {title} {\bibinfo {title} {Deep convolutional neural networks for
  image classification: {A} comprehensive review},\ }\href@noop {} {\bibfield
  {journal} {\bibinfo  {journal} {Neural Comput.}\ }\textbf {\bibinfo {volume}
  {29}},\ \bibinfo {pages} {2352} (\bibinfo {year} {2017})}\BibitemShut
  {NoStop}%
\bibitem [{\citenamefont {Vaswani}\ \emph {et~al.}(2017)\citenamefont
  {Vaswani}, \citenamefont {Shazeer}, \citenamefont {Parmar}, \citenamefont
  {Uszkoreit}, \citenamefont {Jones}, \citenamefont {Gomez}, \citenamefont
  {Kaiser},\ and\ \citenamefont {Polosukhin}}]{transformer}%
  \BibitemOpen
  \bibfield  {author} {\bibinfo {author} {\bibfnamefont {A.}~\bibnamefont
  {Vaswani}}, \bibinfo {author} {\bibfnamefont {N.}~\bibnamefont {Shazeer}},
  \bibinfo {author} {\bibfnamefont {N.}~\bibnamefont {Parmar}}, \bibinfo
  {author} {\bibfnamefont {J.}~\bibnamefont {Uszkoreit}}, \bibinfo {author}
  {\bibfnamefont {L.}~\bibnamefont {Jones}}, \bibinfo {author} {\bibfnamefont
  {A.}~\bibnamefont {Gomez}}, \bibinfo {author} {\bibfnamefont
  {{\L}.}~\bibnamefont {Kaiser}},\ and\ \bibinfo {author} {\bibfnamefont
  {I.}~\bibnamefont {Polosukhin}},\ }\bibfield  {title} {\bibinfo {title}
  {Attention is all you need},\ }\href@noop {} {\bibfield  {journal} {\bibinfo
  {journal} {Adv.\ Neur.\ Inf.\ Proc.\ Syst.}\ }\textbf {\bibinfo {volume}
  {30}} (\bibinfo {year} {2017})}\BibitemShut {NoStop}%
\bibitem [{\citenamefont {Bahdanau}\ \emph {et~al.}(2014)\citenamefont
  {Bahdanau}, \citenamefont {Cho},\ and\ \citenamefont {Bengio}}]{attention}%
  \BibitemOpen
  \bibfield  {author} {\bibinfo {author} {\bibfnamefont {D.}~\bibnamefont
  {Bahdanau}}, \bibinfo {author} {\bibfnamefont {K.}~\bibnamefont {Cho}},\ and\
  \bibinfo {author} {\bibfnamefont {Y.}~\bibnamefont {Bengio}},\ }\bibfield
  {title} {\bibinfo {title} {Neural machine translation by jointly learning to
  align and translate},\ }\href@noop {} {\bibfield  {journal} {\bibinfo
  {journal} {arXiv preprint arXiv:1409.0473}\ } (\bibinfo {year}
  {2014})}\BibitemShut {NoStop}%
\bibitem [{\citenamefont {Brown}\ \emph {et~al.}(2020)\citenamefont {Brown},
  \citenamefont {Mann}, \citenamefont {Ryder}, \citenamefont {Subbiah},
  \citenamefont {Kaplan}, \citenamefont {Dhariwal}, \citenamefont
  {Neelakantan}, \citenamefont {Shyam}, \citenamefont {Sastry}, \citenamefont
  {Askell} \emph {et~al.}}]{GPT3}%
  \BibitemOpen
  \bibfield  {author} {\bibinfo {author} {\bibfnamefont {T.}~\bibnamefont
  {Brown}}, \bibinfo {author} {\bibfnamefont {B.}~\bibnamefont {Mann}},
  \bibinfo {author} {\bibfnamefont {N.}~\bibnamefont {Ryder}}, \bibinfo
  {author} {\bibfnamefont {M.}~\bibnamefont {Subbiah}}, \bibinfo {author}
  {\bibfnamefont {J.}~\bibnamefont {Kaplan}}, \bibinfo {author} {\bibfnamefont
  {P.}~\bibnamefont {Dhariwal}}, \bibinfo {author} {\bibfnamefont
  {A.}~\bibnamefont {Neelakantan}}, \bibinfo {author} {\bibfnamefont
  {P.}~\bibnamefont {Shyam}}, \bibinfo {author} {\bibfnamefont
  {G.}~\bibnamefont {Sastry}}, \bibinfo {author} {\bibfnamefont
  {A.}~\bibnamefont {Askell}}, \emph {et~al.},\ }\bibfield  {title} {\bibinfo
  {title} {Language models are few-shot learners},\ }\href@noop {} {\bibfield
  {journal} {\bibinfo  {journal} {Adv.\ Neur.\ Inf.\ Proc.\ Syst.}\ }\textbf
  {\bibinfo {volume} {33}},\ \bibinfo {pages} {1877} (\bibinfo {year}
  {2020})}\BibitemShut {NoStop}%
\bibitem [{\citenamefont {Pereira}\ \emph {et~al.}(1994)\citenamefont
  {Pereira}, \citenamefont {Tishby},\ and\ \citenamefont
  {Lee}}]{pereira1994distributional}%
  \BibitemOpen
  \bibfield  {author} {\bibinfo {author} {\bibfnamefont {F.}~\bibnamefont
  {Pereira}}, \bibinfo {author} {\bibfnamefont {N.}~\bibnamefont {Tishby}},\
  and\ \bibinfo {author} {\bibfnamefont {L.}~\bibnamefont {Lee}},\ }\bibfield
  {title} {\bibinfo {title} {Distributional clustering of {E}nglish words},\
  }\href@noop {} {\bibfield  {journal} {\bibinfo  {journal} {arXiv preprint
  cmp-lg/9408011}\ } (\bibinfo {year} {1994})}\BibitemShut {NoStop}%
\end{thebibliography}%

\end{document}